\documentstyle[aps,pra]{revtex}
\begin{document}
\title{Interacting Fermi gas in a harmonic trap}
\author{G. M. Bruun and K.\ Burnett}
\address{Department of Physics,
Clarendon Laboratory,
University of Oxford,
Oxford OX1 3PU,
England}
\date{\today}
\maketitle

\begin{abstract}
 In view of ongoing experiments to trap ultracold spin-polarized $^6$Li,
we study various properties of an interacting Fermi gas in a harmonic trap taking 
the  discrete nature of the unperturbed  harmonic trap levels into account exactly.
As $^6$Li 
 has a rather large and negative scattering length, we focus on the effects of the attractive 
atom-atom 
interaction on several thermodynamic properties and on the momentum and density distributions.
 The dependence of the chemical potential, the specific heat and the density and momentum 
distributions on  the number of particles in the trap is obtained. We also calculate the energy of the
 gas. Comparison is made with results  of a  
semiclassical calculation and with the properties of a non-interacting 
gas. We find that the effect of the interactions is rather large for realistic trap frequencies. 
Hence, it is important to include these interactions in any quantitative predictions relevant for 
experiments.
\end{abstract}
\section{Introduction}
Recently there has been a lot of interest in the properties of trapped 
ultra-cold atoms. This interest has largely been sparked by the achievement 
of Bose-Einstein condensation in the Bosonic systems $^{87}$Rb, $^7$Li, and 
$^{23}$Na~\cite{Anderson,Bradley,Davis}.  Several theoretical 
studies of trapped degenerate Fermi gases 
have been presented in addition to those on Bose systems. It has even been shown
 that a two-component gas of 
spin-polarized atomic $^6$Li becomes superfluid at experimentally obtainable 
densities and temperatures~\cite{Stoof}. This is due to the fact that $^6$Li has an 
anomalously large and negative $s$-wave scattering length $a$, with a  recent measurement 
giving  $a\simeq-2160a_0$ where $a_0$ is the Bohr radius~\cite{Abraham}.
$s$-wave scattering is forbidden for two fermions in the same spin-state, 
but, the  $^6$Li atom has six 
hyperfine states. This means that  by trapping $^6$Li in two different hyperfine states, one can 
observe the relatively strong interactions due to  $s$-wave scattering between atoms in different 
 states. The trapping can be achieved since the energy of the hyperfine states depend on the 
external magnetic field. The highest three states ``prefer'' a lower magnetic field and can therefore 
be trapped in a static magnetic trap. In particular, one can trap the two highest states  
as proposed by Houbiers \textit{et al}~\cite{Houbiers}.
These two states will here be labeled $|\uparrow\rangle$ and 
 $|\downarrow\rangle$. In such a two-component gas of trapped spin-polarized $^6$Li atoms, 
there will only be, to a good approximation,  interactions between atoms in different hyperfine 
states, whereas there will be almost no interaction between atoms in the same state. 
Indeed, it was for such a two-component system that the relatively high transition temperature 
 $T_c$ for a BCS-type phase transition was predicted~\cite{Stoof}.

The purpose of this paper is to examine the normal state properties of such a trapped gas 
of $^6$Li atoms with components in the two highest hyperfine states. 
It is clearly necessary  to understand these 
properties in the path to achieving the predicted BCS-type transition. 
The normal state properties of a non-interacting trapped cloud of fermions have been 
treated within the semiclassical Thomas-Fermi approximation~\cite{Butts} and also 
by taking the discrete nature of the trap energy levels in the non-interacting limit 
  into account~\cite{Schneider}.
As the interactions between $^6$Li atoms in two  different hyperfine states are relatively 
strong, it is important to include the effect of these interactions in any realistic treatment 
of the system. Hence, the present paper extends the analysis of Refs.\ \cite{Butts,Schneider}
by including both the discrete nature of the trap levels  as well as  the effects 
 of the  interactions (in the mean-field approximation). 

The paper is organized as follows. In Sec.\ \ref{Formalism} we set up the formalism 
needed to treat an interacting  two-component system of fermions within the mean-field
approximation. We then, in Sec.\ \ref{chemical}, analyze the influence of the trap potential 
and the interactions on the chemical potential of the gas. We show, that the interactions have
two distinct effects on the chemical potential: they lower its value below the non-interacting 
value and  smooth out the steplike features predicted by Schneider 
and Wallis~\cite{Schneider}. The quasiparticle energy spectra and wave functions
are considered in Sec.\ \ref{spectrum} and in Sec.\ \ref{Heatcap} we investigate the
 behaviour of the energy and the 
heat capacity. We find that at sufficiently low temperatures the interactions 
 change the qualitative behaviour of the  heat capacity.
This is  explained in terms of the effect  the interactions have on the quasiparticle spectrum.
In Sec.\ \ref{dkdist} we discuss the density and momentum distribution and their 
deviations from the non-interacting case and from the Thomas-Fermi predictions. 
Finally, we summarize the results in  Sec.\ \ref{conclu}.

\section{Formalism} \label{Formalism}
We consider a dilute gas of interacting  $^6$Li atoms in two hyperfine states trapped in an external 
potential $U_0({\mathbf{r}})$. As the gas is dilute the 
interactions mainly happen through two-body collisions. Furthermore, since the s-wave scattering 
length is much larger than the p-wave scattering length we can neglect any interaction between 
fermions in the same hyperfine state. The gas is then described by the Hamiltonian:
\begin{equation}
\hat{H}=\sum_{\sigma}\int d^3r\,
 \psi_{\sigma}^{\dagger}({\mathbf{r}})
[ \frac{-\hbar^2}{2m}\nabla^2+\frac{1}{2}m\omega^2r^2-\mu]\psi _{\sigma}({\mathbf{r}}) 
-g\int d^3r\,  \psi_{\uparrow}^{\dagger}({\mathbf{r}})\psi_{\downarrow}^{\dagger}
({\mathbf{r}})\psi_{\downarrow}({\mathbf{r}})\psi_{\uparrow}({\mathbf{r}}),
\end{equation}
where  $m$ is the mass of the particles, the attractive interparticle potential has been 
approximated 
by a contact potential $V({\mathbf{r}}'-{\mathbf{r}})\simeq-g\delta({\mathbf{r}}'-{\mathbf{r}})$,
 $g>0$.  The field operators  $\psi_{\sigma}({\mathbf{r}})$ obey 
the usual fermionic anticommutation rules and describe the annihilation of a fermion 
at position ${\mathbf{r}}$ in the hyperfine state $|\sigma\rangle$.  The trapping 
potential is for simplicity assumed to be well described by an isotropic harmonic oscillator 
 $U_0({\mathbf{r}})=\frac{1}{2}m\omega^2 r^2$. The trapping frequency $\omega$  is
 taken to be the same for each hyperfine state. We have  
assumed that number of particles $N$ in each state is the same such that we
 only have one chemical potential $\mu$. As the critical temperature for a BCS type 
transition is maximum when the 
number of particles in the two hyperfine levels is equal~\cite{Stoof,Houbiers}, we expect this 
configuration to have the most experimental relevance. 
The non-interacting case is achieved by setting $g=0$; this limit has been treated 
 by Butts and Rokhsar within the Thomas-Fermi approximation~\cite{Butts} and by Schneider 
and Wallis~\cite{Schneider} taking the quantizing effect of the trap potential into account.
In this paper, we are interested in the effect of the interactions on the normal state properties
 of the gas. We can, therefore, ignore any pairing 
correlations leading to a BCS type transition and use the following  mean-field Hamiltonian:
\begin{equation}
\hat{H}_{Mean}=\sum_{\sigma}\int d^3r\,
 \psi_{\sigma}^{\dagger}({\mathbf{r}})
[ \frac{-\hbar^2}{2m}\nabla^2+\frac{1}{2}m\omega^2r^2+
U({\mathbf{r}})-\mu]\psi _{\sigma}({\mathbf{r}}).
\end{equation}
Here the self-consistent field 
 $U({\mathbf{r}})=-g\langle\psi_{-\sigma}
 ^{\dagger}({\mathbf{r}})\psi_{-\sigma}({\mathbf{r}})\rangle$
 is the standard Hartree-Fock result for a contact interaction. To include all 
two-body scattering processes on the mean-field level, one can put 
 $g=4\pi|a|\hbar^2/m$, with $a$ being the s-wave
 scattering length of collisions between the fermions in the two hyperfine states~\cite{Stoof2}.
As the trapping potential is 
assumed to be isotropic, the  self-consistent solution of  the mean-field 
Hamiltonian with the lowest energy will be spherically symmetric. The Hamiltonian can 
then readily be diagonalized by writing the field operator as 
\mbox{$
\psi_{\sigma}({\mathbf{r}})=\sum_{\nu lm}u^{\nu}_{lm}({\mathbf{r}})a^{\nu}_{\sigma lm}
$ }.
The operator $a^{\nu}_{\sigma lm}$ describes the annihilation of a quasiparticle in 
the hyperfine state $\sigma$ with total angular momentum $[l(l+1)\hbar^2]^{1/2}$, a component 
along an arbitrary z-direction of $m\hbar$ and a wave function $u^{\nu}_{lm}({\mathbf{r}})$.
 We write  the quasiparticle wave function in the form
\begin{equation} \label{qpw}
u^{\nu}_{lm}({\mathbf{r}})=\frac{u^{\nu}_l(r)}{r}Y_{lm}(\theta,\phi),
\end{equation}
where $Y_{lm}(\theta,\phi)$ are the usual orbital angular momentum eigenfunctions.
 $\hat{H}_{Mean}$ is then diagonalized by solving
\begin{equation} \label{HFeqn}
E^{\nu}_lu^{\nu}_l(r)=[-\frac{\hbar^2\partial_r^2}{2m}+\frac{\hbar^2l(l+1)}{2mr^2}
+\frac{1}{2}m\omega^2r^2+U(r)-\mu]u^{\nu}_l(r)
\end{equation}
for each angular momentum $l$.
Here the quasiparticle energy $E^{\nu}_l$ is independent of $m$ due to the spherical 
symmetry. The self-consistent potential is determined by
\begin{equation}\label{U(r)}
U(r)=-g\sum_{\nu lm}|u^{\nu}_{lm}({\mathbf{r}})|^2f(E^{\nu}_l)=-g\sum_{\nu l}
\frac{u^{\nu}_l(r)^2}{r^2}\frac{2l+1}{4\pi}f(E^{\nu}_l)
\end{equation}
where $f(x)=1/[\exp(x/k_BT)+1]$ is the Fermi function and the addition theorem for spherical 
harmonics has been used. Numerically, we use a cutoff of the order  
 $E^{\nu}_l\:\raisebox{-0.4ex}{$\stackrel {<}{\sim}$}\: 2\mu$ for the sum 
given in Eq.\ (\ref{U(r)}). This is more than sufficient since we in this paper are considering low 
temperatures  ($k_BT<\hbar\omega$) and  we know that  levels with energies higher than a few 
 $k_BT$ give a vanishing contribution to the density.
 Thus, within the mean-field approximation, the problem of an
 interacting gas of fermions in two hyperfine states in an isotropic harmonic trap is
 equivalent to solving Eq.\ (\ref{HFeqn})-(\ref{U(r)}) self-consistently. Once a solution is 
found, one can easily calculate various observables for the gas. The total energy $E$ for the 
particles in both hyperfine states  is given by 
\begin{eqnarray} \label{energy}
E&=&\langle \hat{H}\rangle+2\mu N\nonumber \\
&=& 2\sum_{\nu l}(2l+1)f(E^{\nu}_l)
\int d^3r\, u^{\nu}_{l0}({\mathbf{r}})( \frac{-\hbar^2}{2m}\nabla^2+
\frac{1}{2}m\omega^2r^2)u^{\nu}_{l0}({\mathbf{r}})-\frac{1}{g}\int d^3rU(r)^2.
\end{eqnarray}
Here we have added $2N$ to $\langle \hat{H}\rangle$ as it is the total number of
 particles in the trap.
In the following sections we will solve the above equations for  various sets of parameters and 
 calculate several observables from the solutions.

 As the negative scattering length introduces 
an attractive interatomic potential, the system can collapse to a fluid or solid state when the 
density of particles becomes too large. This effect has been examined by Houbiers 
\textit{et al.}~\cite{Houbiers} within the semiclassical Thomas-Fermi approximation. 
For a trap with an equal number of particles in each hyperfine state they find, that the 
spinoidal point is given by $N^{1/6}|a|/l\simeq 0.66$  with 
$l=\sqrt{\hbar/\omega m}$ being the trap length. Numerically, this transition is seen from the 
fact that there is no self-consistent solution to Eq.\ (\ref{HFeqn})-(\ref{U(r)}) when the 
chemical potential is too high for a given coupling strength. The density of particles increases 
for each iteration without bound indicating that the system is collapsing into a new dense phase. 
We find that this problem arises in the region of parameters where $N^{1/6}|a|/l\sim O(1)$ in 
qualitative agreement with Houbiers \textit{et al.} However, due to the computation load we have
 not been able to verify in detail the prediction  $N^{1/6}|a|/l\simeq 0.66$ 
for the spinoidal transition line. 
As we  do not have an appropriate theory for such a phase we will in this paper work with 
parameters ($N,g$) such that we are well below this spinoidal phase-transition line. 

We shall compare some of our results obtained from the solution of  
Eq.\ (\ref{HFeqn})-(\ref{U(r)}) with approximate results based on the 
Thomas-Fermi approximation.
This approximation essentially treats the trap potential as being locally 
constant. The Hartree-Fock equations are then trivially solved for 
each ${\mathbf{r}}$ by plane waves. The quasiparticle energies are given by
\begin{equation}\label{QPETF}
E_k(r)=\frac{\hbar^2k^2}{2m}-g\rho(r)-\mu(r).
\end{equation}
Here the local chemical potential is $\mu(r)\equiv \mu-\frac{1}{2}m\omega^2 r^2$ and 
 the density is given by 
\begin{equation}\label{DensTF}
\rho(r)=\int \frac{d^3k}{(2\pi)^3}f[E_k(r)].
\end{equation}
Eq.\ (\ref{QPETF})-(\ref{DensTF}) has to be solved self-consistently at each point 
 ${\mathbf{r}}$. The total number of particles in a single hyperfine state in the system is then 
 $N=\int d^3r\, \rho(r)$. 
 
\section{The chemical potential} \label{chemical}
In this section we will determine the chemical potential $\mu$ as a function of the number of 
particles in the trap. Schneider and Wallis~\cite{Schneider} 
have done an extensive analysis of $\mu(N)$ in the case of a non-interacting 
gas. They found some remarkable steplike features in $\mu(N)$ as compared to the 
Thomas-Fermi approximation; these steps were  due to the shell structure of 
 the energy spectrum of an isotropic harmonic trap. They did, however, predict that for 
a \textit{real} 
Fermi gas, the interactions would tend to smooth out the steps. Using the formalism outlined
 above, we are now able to examine in detail how the interactions affect these steplike features. In 
Fig.\ (1) 
we have plotted the chemical potential $\mu$ as a function of the number of 
particles $N$ 
in a single hyperfine state for various coupling strengths $g$ for a very low temperature 
($k_BT\ll \hbar\omega$). These 
curves were obtained by solving  Eq.\ (\ref{HFeqn})-(\ref{U(r)}) for varying
  $\mu$ and $g$.
The dashed lines correspond to the Thomas-Fermi approximation obtained from the solution 
of Eq.\ (\ref{QPETF})-(\ref{DensTF}). 
In Fig.\ (1) (a) we have compared  $\mu(N)$ 
for the rather large coupling strengths $g'=1$ and $g'=2$ with the $g=0$ case 
when there are relatively few particles in the trap in order to highlight the steplike features.
 We have defined $g'\equiv g/(\hbar\omega l^3)$. As can be seen, that the interaction has
 two effects. First of all, it lowers the value 
of the chemical potential for a given number of particles $N$ as compared to the 
 $g=0$ case. This is as expected, since the mutual attraction between the particles in the two 
hyperfine states lowers the energy of the gas. Secondly, we note that the step like features of 
 $\mu(N)$ are smoothed out by the interaction. For $g'=1$ the
 step like features survive up to  $N\,\raisebox{-0.4ex}{$\stackrel {<}{\sim}$}\,2000$ 
and for $g'=2$
they survive for $N\,\raisebox{-0.4ex}{$\stackrel {<}{\sim}$}\,500$. Fig.\ (1) (b)
 show $\mu(N)$ for $g'=0.4$ and $g'=0.2$. Putting 
 $|g|=4\pi |a|\hbar^2/m$, with $a=-2160a_0$ being the s-wave scattering length for $^6$Li,  
 $g'=0.4$ corresponds to a trap frequency of $\nu=\omega/2\pi=144$Hz 
and $g'=0.2$ to a trap frequency of $\nu=\omega/2\pi=288$Hz. These 
values  are close to the ones used in the Bose-Einstein condensation  
experiments at JILA~\cite{JILA} and at 
MIT~\cite{MIT} respectively. We again see the same behaviour as in Fig.\ (1) (a). 
The step like features break down for $N\approx 4\times10^4$ for 
 $g'=0.4$. For $g'=0.2$ they break down at $N\approx 10^5$ which is not shown 
in Fig.\ (1). 
 
\section{The energy spectrum} \label{spectrum}
From Fig\ (1) we see, that the step like features survive even though the value 
of $\mu(N)$ is several $\hbar\omega$ lower than the $g=0$ prediction. This is somewhat 
surprising as the steps are associated with the gaps of size $\hbar\omega$ in the $g=0$ spectrum. 
One might expect these gaps to disappear once the levels are lowered by more 
than $\hbar\omega$
due to the interactions. To examine this effect, we have in Fig.\ (2) 
 plotted the quasiparticle spectrum for $g'=2$ 
for a very low temperature for each angular momentum $l$. In Fig.\ (2)  (a) the 
chemical potential is $\mu/\hbar\omega=8$ whereas $\mu/\hbar\omega=13$ in 
Fig.\ (2) (b).
The x-axis denotes the angular momentum $l$ of the quasiparticle states and the quasiparticle 
energies are marked by the symbol $\times$. For comparison the symbols $\circ$ denote the 
non-interacting quasiparticle energies. The chemical potential is indicated by a thin dotted line. 
In the non-interacting 
case ($g=0$), the energy spectrum is given by $E_{\nu}=(\nu+3/2)\hbar\omega$ with 
 $\nu=0, 1, 2 \ldots$. The degeneracy of each level is $D_{\nu}=(\nu+1)(\nu+2)/2$ corresponding
 to the angular momentum $l$ being $l=0,2,\ldots \nu$ for $\nu$ even and 
$l=1,3,\ldots \nu$ for $\nu$ odd. This degeneracy in $l$ gives 
rise to the shell structure of the spectrum which
is indicated in Fig.\ (2) by horizontal dashed lines. In the interacting case, the energy 
 $E^{\nu}_l$ in 
general depends on both $\nu$ and $l$. From Fig.\ (2) (a) ($\mu=8\hbar\omega$) 
we see, that the quasiparticle energies in the interacting case are lowered several 
 $\hbar\omega$ as compared to the normal state energies; e.g. the lowest energy is 
 $-1.7\hbar\omega$ for $g'=2$ as compared to $1.5\hbar\omega$ for $g'=0$. However, 
from the solid lines in Fig.\ (2) (a) we see, that 
the energies still depend only weakly on the angular momentum $l$.
This means that the Hartree potential lowers the energy of the $l=\nu$ state by almost the same 
amount as it lowers the $l=0$ ($\nu$ even) or $l=1$ ($\nu$ odd) state. This is not 
\emph{a priori} 
obvious as the wave functions for those two states are completely different. This is shown in 
 Fig.\ (3) where the solid lines denote the quasiparticle wave functions 
 $u^{\nu=8}_{l=0}(r)$ (a) and $u^{\nu=8}_{l=8}(r)$ (b) for $\mu=8\hbar\omega$ and $g'=2$. These 
states have the energies
 $E^{\nu=8}_{l=0}=8.2\hbar\omega$ and   $E^{\nu=8}_{l=8}=8.4\hbar\omega$ respectively. 
The $l=0$  wave function is spread out from $r=0$ to 
 $r\approx r_{class}=\sqrt{2E_n/m\omega^2}=4l$
 whereas the $l=\nu$ wave function is peaked around $r\approx 2.8l$.
Hence one might expect that the Hartree potential, which is also plotted as a solid line in  
Fig.\ (3) (c), would affect the quasiparticle states 
completely differently. But in fact, the lowering of the two energies as compared to the 
 $g=0$ case is approximately the same. This explains the fact depicted in Fig.\ (1) that 
even though the chemical potential is several $\hbar\omega$ below the $g=0$ prediction,  
 $\mu(N)$ still exhibits the step like features. The reason is, that even though each quasiparticle 
energy is lowered several $\hbar\omega$ due to the interaction, there is still an approximate 
 $l$-degeneracy: there are still bands separated in energy by $\approx \hbar\omega$ 
in the energy spectrum as a function of $l$.
These bands (shell structure) give, as in the $g=0$ case,  rise to the steplike structure
 of $\mu(N)$. 
For comparison, we have also in Fig.\ (3) plotted the non-interacting 
wave functions as dashed lines. As expected, we see that the effect of the interactions is to 
compress the quasiparticle states closer to the center of the trap where the Hartree field is
large. The cloud of particles is compressed as will also be seen from the density distributions 
plotted in Sec.\ (\ref{dkdist}). 
We can qualitatively understand the slight increase in energy with increasing 
 $l$ as seen from the solid lines in Fig.\ (2) (a). The low angular momentum functions 
have a larger amplitude in the center of the trap and are thus more affected by the Hartree
field than the high $l$ states.

Contrary to the $\mu=8\hbar\omega$ case, we see from the solid lines in 
Fig.\ (2) (b), that when $\mu=13\hbar\omega$ 
the approximate independence of the quasiparticle energies on $l$ no longer holds. 
The Hartree field is now so strong that it has washed out the shell structure of the 
spectrum and it is qualitatively different from the $g=0$ case. The energies now increase 
significantly with increasing $l$. This explains the fact
 from Fig.\ (1) (a), that for $\mu\sim13$ and 
 $g'=2$ the steplike features have disappeared. 

\section{The energy and the heat capacity} \label{Heatcap}
For a gas of particles in a constant confining potential, the most useful definition of 
the heat capacity is~\cite{Ensher} is $C_N\equiv\frac{1}{2N} \partial E/\partial T|_N$ with $E(T,N)$ 
given by Eq.\ (\ref{energy}). As pointed out by Schneider and Wallis~\cite{Schneider}, the 
shell structure of the harmonic trap spectrum has a drastic consequence for the low temperature 
($k_BT\ll\hbar\omega$) heat  capacity; the  gaps in the energy spectrum  for $g=0$ mean 
that the heat capacity 
is exponentially suppressed for low temperatures. In the non-interacting case the total energy for 
particles in both hyperfine states is given by
\begin{equation}
E(T)=2\sum_{\nu} E_{\nu} D_{\nu}f(E_{\nu}).
\end{equation}
When the number of particles is such that a finite number of energy levels up to and including the 
level $E_{\nu_F}$ are completely filled for 
 $T=0$ (i.e.\ $N=1,4,10,20,\ldots$ for $\nu_F=0,1,2,3\ldots$),  the low temperature 
chemical potential for a constant number of particles is given by 
\begin{equation}
\mu(T)=E_{\nu_F}+\frac{\hbar\omega}{2}-\frac{1}{2}\ln(\kappa_{\nu_F})k_BT
\end{equation}
with $\kappa_{\nu_F}=D_{\nu_F+1}/D_{\nu_F}>1$. The low temperature heat capacity is then 
easily found to be 
\begin{equation}
\frac{C_{N}(T)}{2Nk_B}=[(\nu_F+5/2)D_{\nu_F+1}\frac{1}{\sqrt{\kappa_{\nu_F}}}
-(\nu_F+3/2)D_{\nu_F}
\sqrt{\kappa_{\nu_F}}]\frac{(\hbar\omega)^2 e^{-\beta\hbar\omega/2}}{2N(k_BT)^2}
\end{equation}
where $\beta=1/k_BT$. This is,  as one would expect, suppressed by a factor 
 $\exp(-\beta\hbar\omega/2)$ as compared to the usual low temperature Thomas-Fermi result:
\begin{equation} \label{TFCN}
 \frac{C_N(T)}{2Nk_B}=\frac{\pi^2k_BT}{\hbar\omega(6N)^{1/3}}.
\end{equation}

 It is interesting to examine how the interactions change this result. 
In Fig.\ (4)  we have plotted the energy per particle $E'/2N$ of the 
interacting gas as a function of temperature $T'\equiv T/\hbar\omega$, where 
 $E'\equiv E/\hbar\omega$ and the energy
 is given by Eq.\ (\ref{energy}).  We have used  the parameters 
  $g'=0.4$, $N=5456$ (a) and $g'=0.4$, $N=43680$ (b). 
For comparison we have also plotted the $g=0$ results as dashed lines in Fig.\ (4)
 (c)-(d).
As can be seen the interactions for both set of parameters have lowered the total energy 
of the gas considerably. This is as expected as we from Fig.\ (1) see, that 
the chemical potential in both cases is significantly lower than the $g=0$ result.
Fig.\ (5)  depicts the corresponding heat capacity as a function
 of the temperature for $N=5456$ (a) and  $N=43680$ (b) (solid lines). 
The dashed lines are the  $g=0$ results and the dash-dotted line is the 
Thomas-Fermi result for $g'=0$ as given by Eq.\ (\ref{TFCN}). 
For $N=5456$ (Fig.\ (5) (a)) we see, that the heat 
capacity for the interating system is still exponentially suppressed for $k_BT\ll\hbar\omega$. 
 For very low $T$ the heat capacity is practically zero as for the $g=0$ case  and the interactions 
have not changed the qualitative behaviour of the heat capacity. 
This is a direct result of the fact that the interactions have not changed the shell structure of the 
energy spectrum. This can be seen from Fig.\ (1) (b)) where the step like 
features of $\mu(N)$ are still prevailing for $g'=0.4$ and $N=5456$. 

The break-down of this effect is shown in Fig.\ (5) (b). 
We see that for $N=43680$ the interactions have now increased heat capacity
 substantially 
over the  non-interacting result for low temperatures. It has the same order of magnitude as
 the Thomas-Fermi prediction and it is qualitatively different from the non-interacting case.
 This is due to the fact that for this set of parameters 
the interactions have washed out the shell structure in the energy spectrum  (the $l$-degeneracy)
  as can be seen from Fig.\ (1), where the steplike features are smoothed out 
for $N=43680$ and $g'=0.4$. There are no gaps of $\hbar\omega$ in the spectrum 
and hence no exponential damping factor in the heat capacity.

It should be noted that for very low temperatures the possibility of a transition to 
a superfluid state arises~\cite{Stoof}. From the well known result of weak coupling 
superconductors, we expect that this transition will give rise to a 
kink in the $E(T)$ curve as the 
energy per particle is lowered as compared to the normal state~\cite{Tinkham}.  
From this kink, we obtain 
a discontinuity in $C_N(T)$ at the transition temperature $T_c$. We will not in this paper 
consider this effect as we are
concentrating on the normal state behaviour. Work is under progress to examine for 
which temperatures and densities this transition takes place and how it can be determined 
experimentally.

 We conclude, that the interactions in general  lower the total energy as one would expect.
The effect of the interaction on the heat capacity depends on the strength of the interactions and on
 the number of particles in the system. When few particles are in the system or when the 
interaction is so weak that the shell structure of the energy spectrum is intact, the heat 
capacity is still exponentially suppressed for low temperatures. The heat capacity can be 
suppressed although the total energy of the system is significantly  lower than for the 
non-interacting case.
 This is, of course, a consequence of the fact that the lowering of the 
quasiparticle energies and the washing out of the shell structure happen at different 
set of parameters $g$ and $N$ as explained in Sec.\ \ref{chemical}-\ref{spectrum}.
Once the interactions are strong 
enough to wash out the shell structure around the chemical potential, the behaviour of the 
heat capacity changes qualitatively. It is no longer suppressed and has the same order of magnitude 
as the Thomas-Fermi prediction. The transition between those two limiting 
behaviours is smooth as a function of $N$ or $g$.

 \section{Density distributions} \label{dkdist}
In this section we will present results for the density and momentum distributions.
 In an isotropic trap, these distributions will be spherically symmetric when the gas 
is in the ground state. The density distribution $\rho(r)$ is calculated from 
 $\rho(r)=-U(r)/g$ with $U(r)$ given by Eq.\ (\ref{U(r)}). In Fig.\ (6), 
 $\rho(r/l)$ is displayed for $N=120, 165$ and $g'=2$. 
  For comparison we have also plotted the distributions for 
 $g'=0$. The dashed curves are  the Thomas-Fermi results obtained from the solution of 
Eq.\ (\ref{QPETF})-(\ref{DensTF}) with a given $N$. The temperature is taken to be zero.
As can be seen, $\rho(r/l)$ is substantially changed due to the interactions. The cloud 
of particles is compressed as compared to the non-interacting result due to the attractive forces. 
As a high density of particles increases the critical temperature for a BCS-type transition, this 
effect favours the formation of the superfluid state~\cite{Stoof}.
Furthermore, we observe  a central minimum for $N=120$  and a central 
peak for $N=165$ for both $g'=2$ and $g'=0$ as compared to the Thomas-Fermi 
predictions. For the $g'=0$, this is because 
 $N=165$ corresponds to a filled shell with $\nu_F=8$,  
 $N=120$ corresponds to a filled shell with $\nu_F=7$ and the fact that 
 shells with $\nu=$ odd do not contribute to $\rho(0)$ as they have odd angular 
momentum~\cite{Schneider}. We see that the interacting system exhibit the same 
qualitative dependence of $\rho(0)$ for this set of parameters, even though the actual 
densities are substationally different from their $g'=0$ counterparts. This is due to the fact, that 
for this set of parameters, the shell structure of the  quasiparticle spectrum is still intact although 
the actual energies and wave functions are changed substantially. This can be 
seen from Fig.\ (1) (a) where the step like features still prevail for this set of parameters.
 Thus, even for $g'=2$, $N=165$ still corresponds a  highest shell of even angular momentum states 
totally filled whereas  $N=120$ corresponds a filled highest shell of odd angular 
momentum states. For a larger number of particles, where the shell structure of the quasiparticle 
spectrum has been washed out, it turns out that 
this behaviour of $\rho(0)$ has disappeared as expected. Also, a non-vanishing temperature
will tend to wash out the predicted dip/top behaviour of the density as the transition from occupied to 
unoccupied shells as a function of energy becomes less abrupt.

The momentum distribution of the particles is not completely straightforward to measure. A simple
free expansion experiment where one switches off the trap potential non-adiabatically does 
not strictly measure this distribution. The reason is, that the gas does not expand freely; collisions 
between particles in different hyperfine states will alter the momentum distribution. One would 
need a time-dependent formalism~\cite{Holland} in order to treat such an expansion 
rigorously. However,  we will here assume that the momentum 
distribution in the trapped state gives a good indication of the distribution measured 
in such a free expansion experiment.
 The spherically symmetric momentum distribution 
 $\langle c^\dagger_kc_k\rangle$, where  $c^\dagger_k$ creates a particle 
in a plane wave state $\exp(ikz)$ along an arbitrary z-direction, is calculated from 
\begin{equation}
\langle c^\dagger_kc_k\rangle=\sum_{\nu l}|\langle k|u^{\nu}_{l0}\rangle|^2
f(E^{\nu}_l).
\end{equation}
Here we have utilized the fact, that a plane wave along the $z-$direction only contains 
 $m=0$ spherical harmonics.
Using the well known expansion of a plane wave $\exp(ikz)$ in spherical harmonics~\cite{Schiff} we 
obtain
\begin{eqnarray}
\langle k|u^{\nu}_{l0}\rangle&=&\int d^3r\,e^{-ikz}u^{\nu}_{l0}({\mathbf{r}})\nonumber \\
&=&(-i)^l\sqrt{(2l+1)4\pi}\left(\frac{\pi}{2k}\right)^{1/2}
\int_0^\infty J_{l+1/2}(kr)u^{\nu}_l(r)\sqrt{r} dr
\end{eqnarray}
where $J_{l+1/2}(x)$ is the ordinary Bessel function. In Fig.\ (7), we have 
plotted the momentum distribution for 
 $N=43680$ and $g'=0.4$. The momentum is measured in units of $l^{-1}$. 
The dashed curve is  the non-interacting result.
For this high number of particles, the shell structure of the quasiparticle spectrum is washed out and 
the  distributions are, apart from a small shell  around the edge of the cloud, almost 
identical to the Thomas-Fermi prediction (not plotted).
We see that the interactions have spread out the distribution considerably as compared to the 
non-interacting case. As the Hartree field lowers the quasiparticle energies more levels 
become occupied in the center of the trap. The higher momentum states thus become populated 
 leading to a spreading out of the momentum distribution as compared to the $g'=0$ case.

We conclude that the interactions in general alter both the momentum distribution and the 
density distribution substantially. The density distribution is compressed and the 
momentum distribution spread out as compared to the non-interacting results. Furthermore, 
the central minima and maxima of $\rho(r)$ as a function of $N$ can still be observed, when 
the interactions have not yet washed out the shell structure of the quasiparticle spectrum.

\section{Conclusion} \label{conclu}
In this paper we have considered a trapped spin-polarized gas of interacting fermions. 
We find that the interactions have two distinct effects on the quasiparticle spectrum. 
It lowers the quasiparticle energies and thus the total energy of the gas. Also, above 
a certain number of particles in the trap, it washes 
out the shell like structure of the spectrum associated with a harmonic trap. These two 
effects are independent, in the sense that the energy of the gas can be lowered considerably 
as compared to the non-interacting case, but the shell structure of the spectrum is left
relatively intact. One can still, therefore, observe effects such as step like features in the 
chemical potential, exponential damping of the low temperature heat capacity and 
maxima and minima in  $\rho(0)$ associated with a non-interacting gas. 
Whether these effects will be observable depend on whether the condition 
 $k_BT\ll \hbar\omega$ is experimentally feasible.
The interactions are also found to compress 
the atom cloud and spread out the momentum distribution considerably. This effect should be 
readily observable; it is important to include in any realistic calculation of the properties 
of spin-polarized $^6$Li in a trap.

\section{Acknowledgments}
 This work was supported  by the Engineering and Physical Sciences Research Council.
 We should also like to acknowledge valuable discussions with R.\ Dum.

\newpage
\begin{center}{Figure Captions}\end{center}
\bigskip
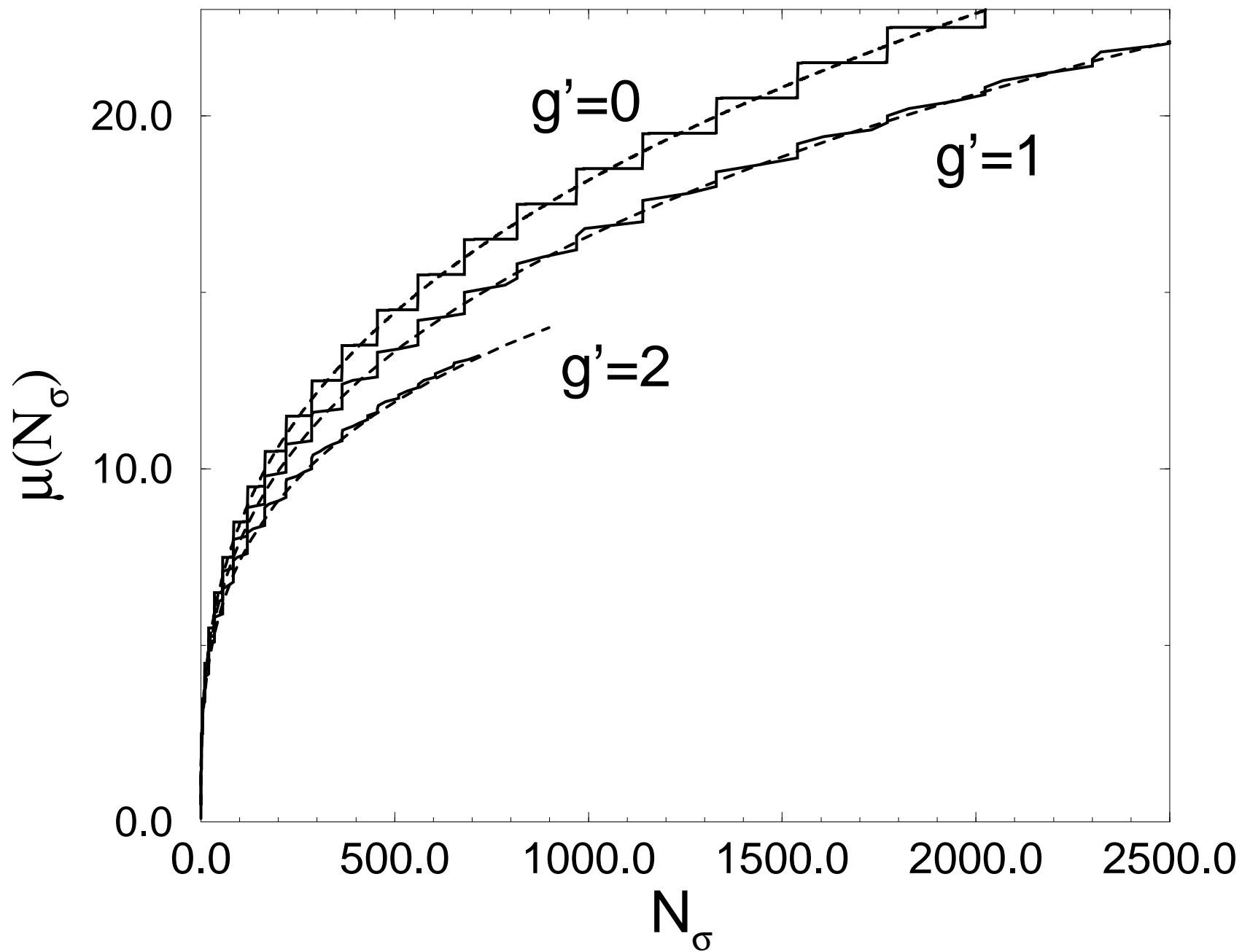
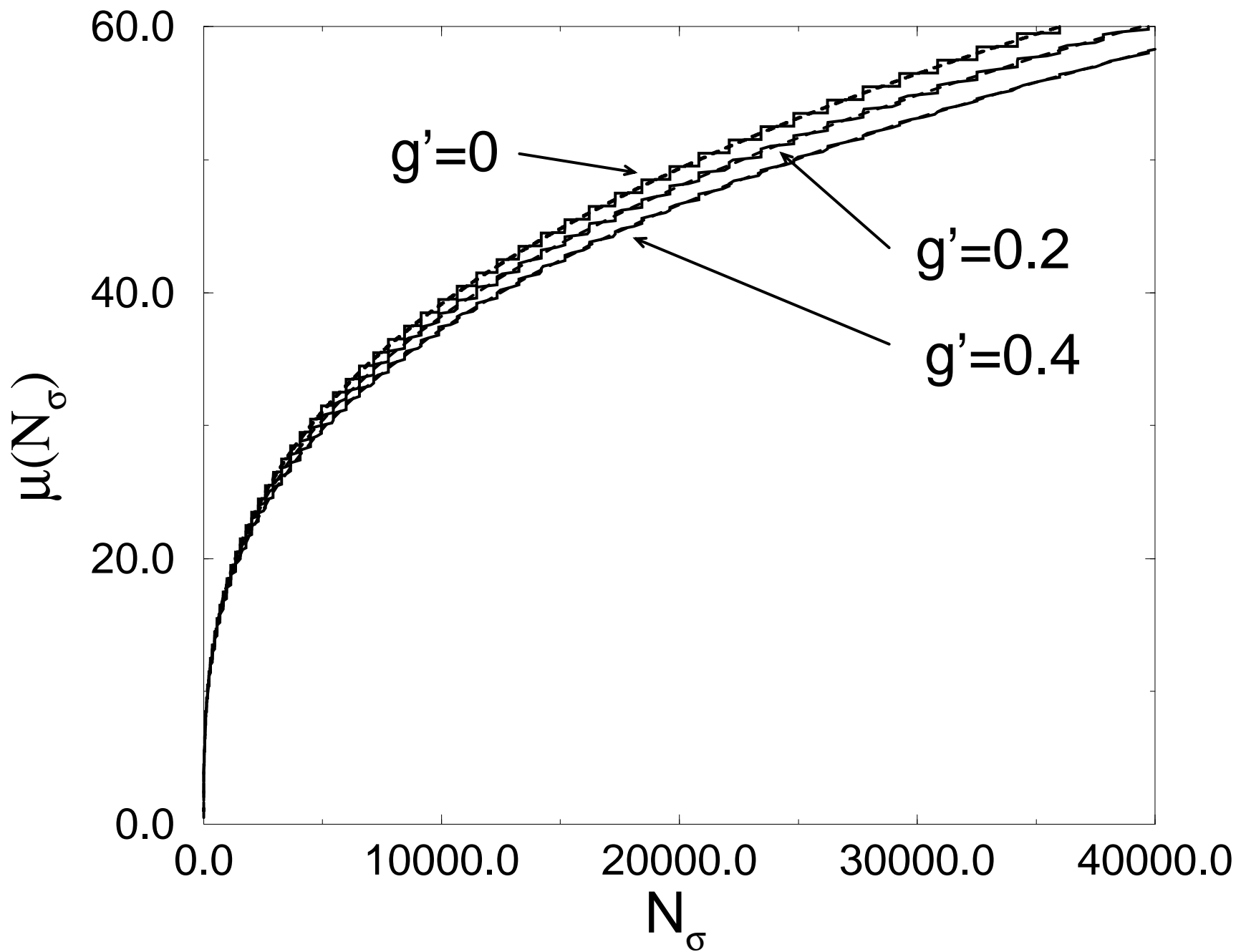
\noindent Fig.\ 1: $\mu(N_{\sigma})$ in units of $\hbar\omega$ for $g'=1$, $g'=2$ (a) and 
 $g'=0.2$, $g'=0.4$ (b). The dashed lines are the Thomas-Fermi approximation.

\

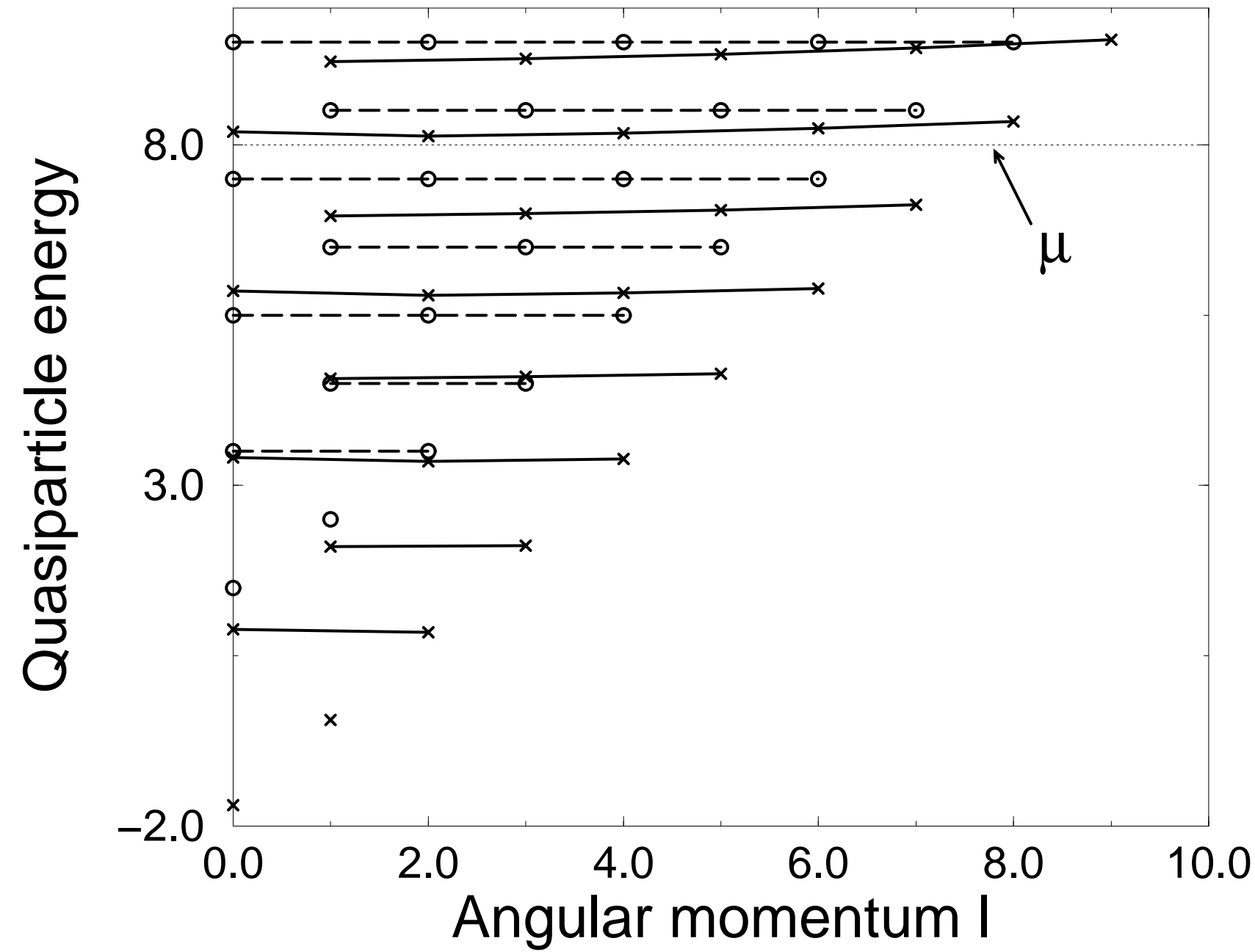
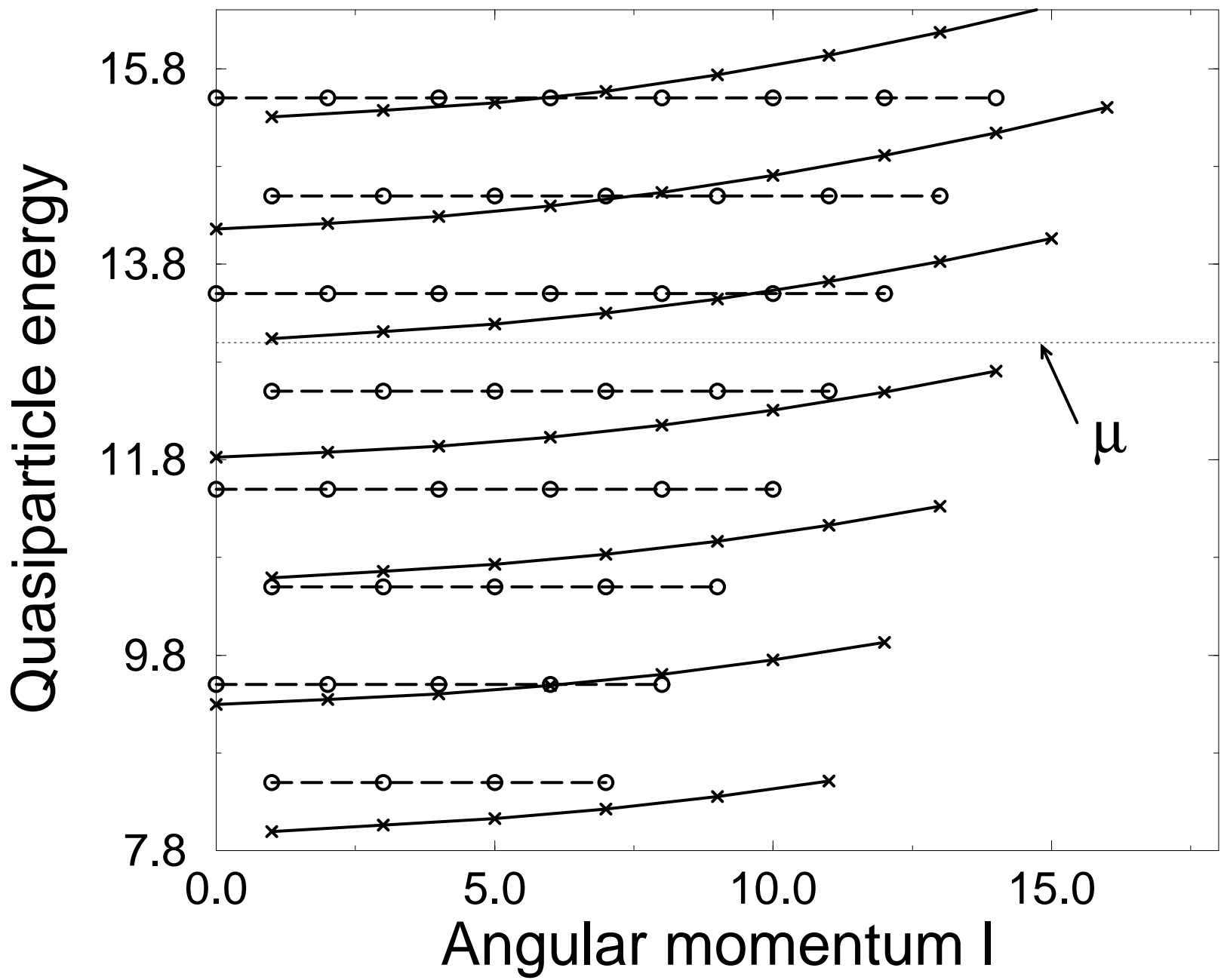
\noindent Fig.\ 2: The quasiparticle spectrum in units of $\hbar\omega$ 
for $g'=2$ , $\mu/\hbar\omega=8$ (a)  and
 $g'=2$, $\mu/\hbar\omega=13$ (b).

\
 
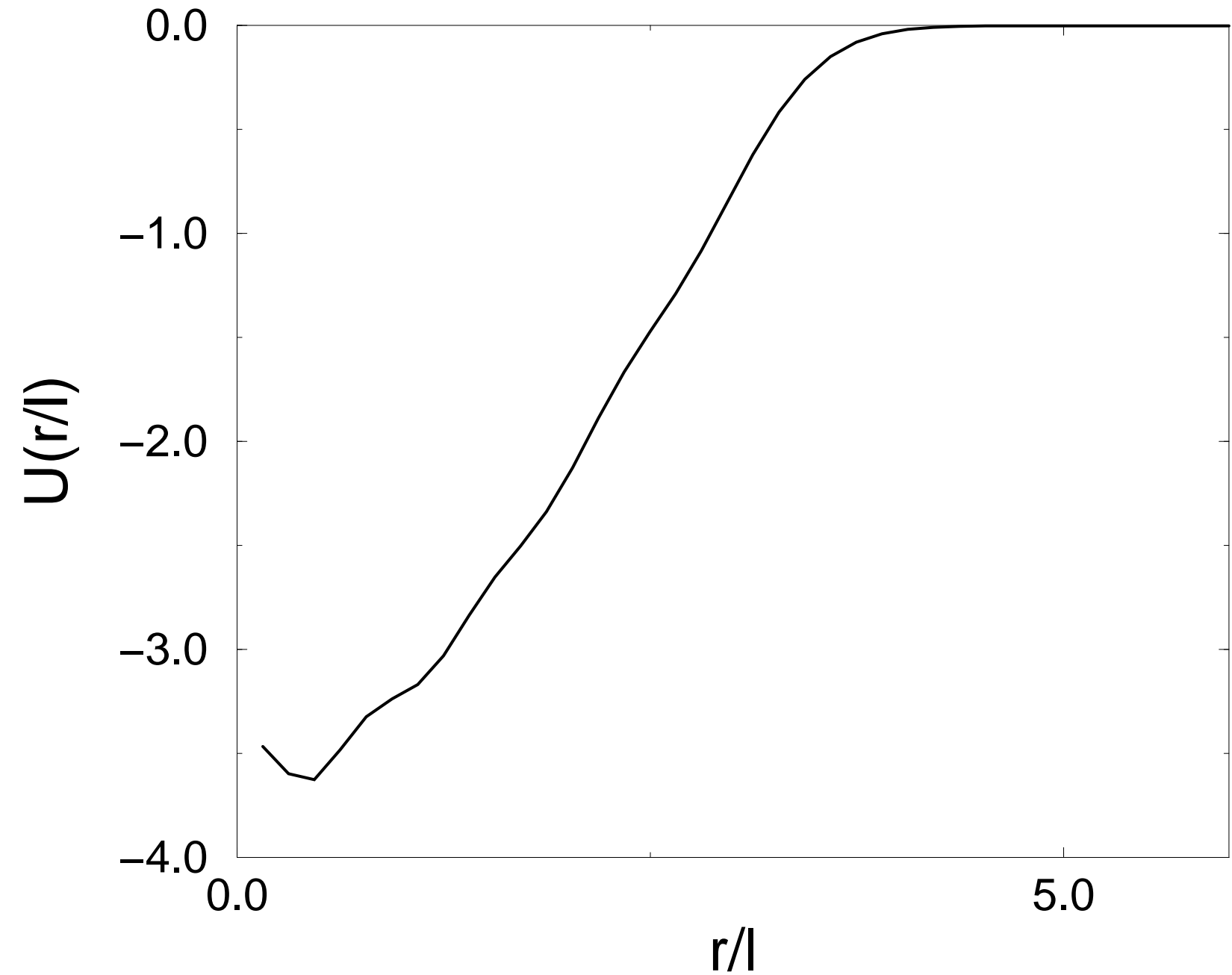
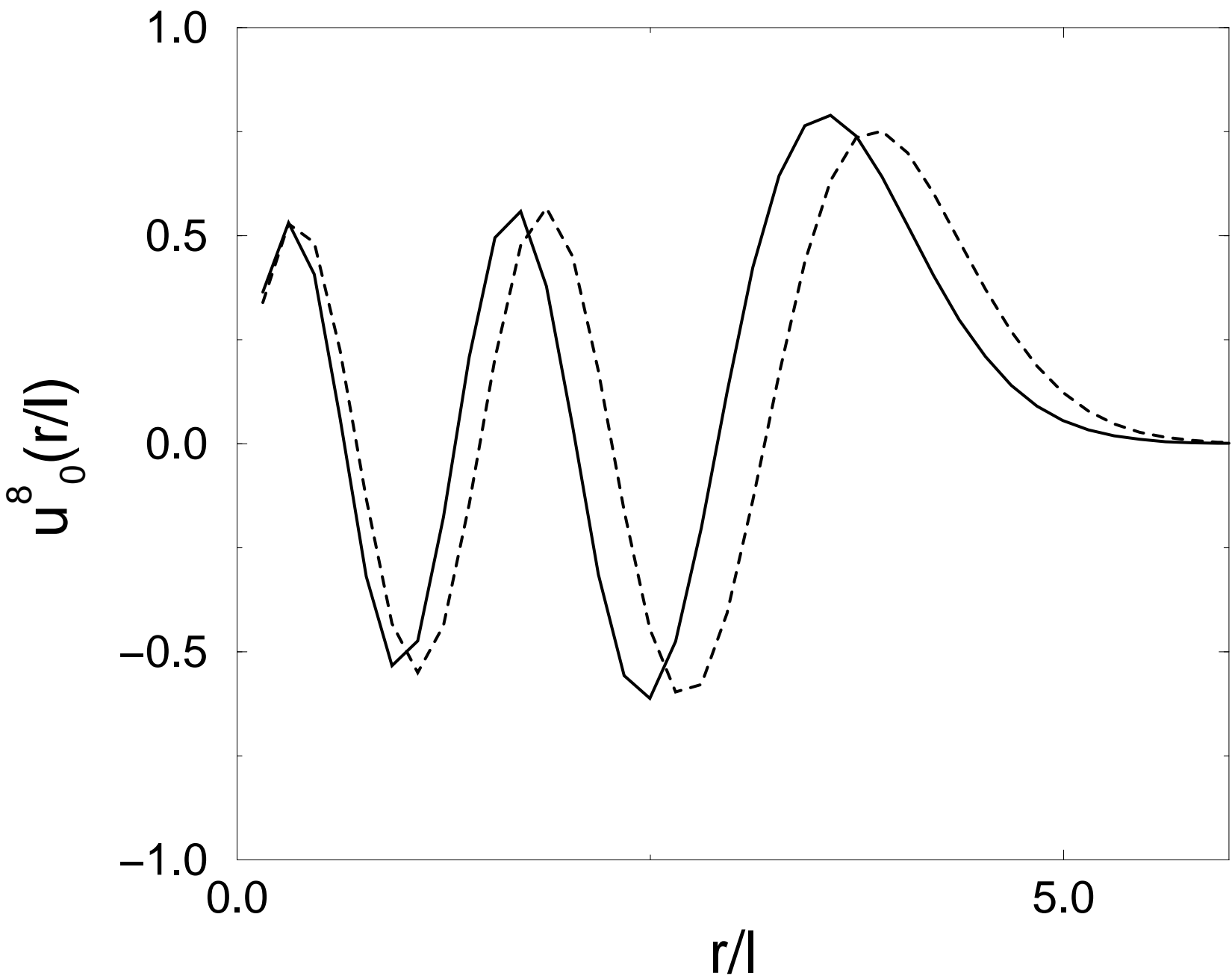
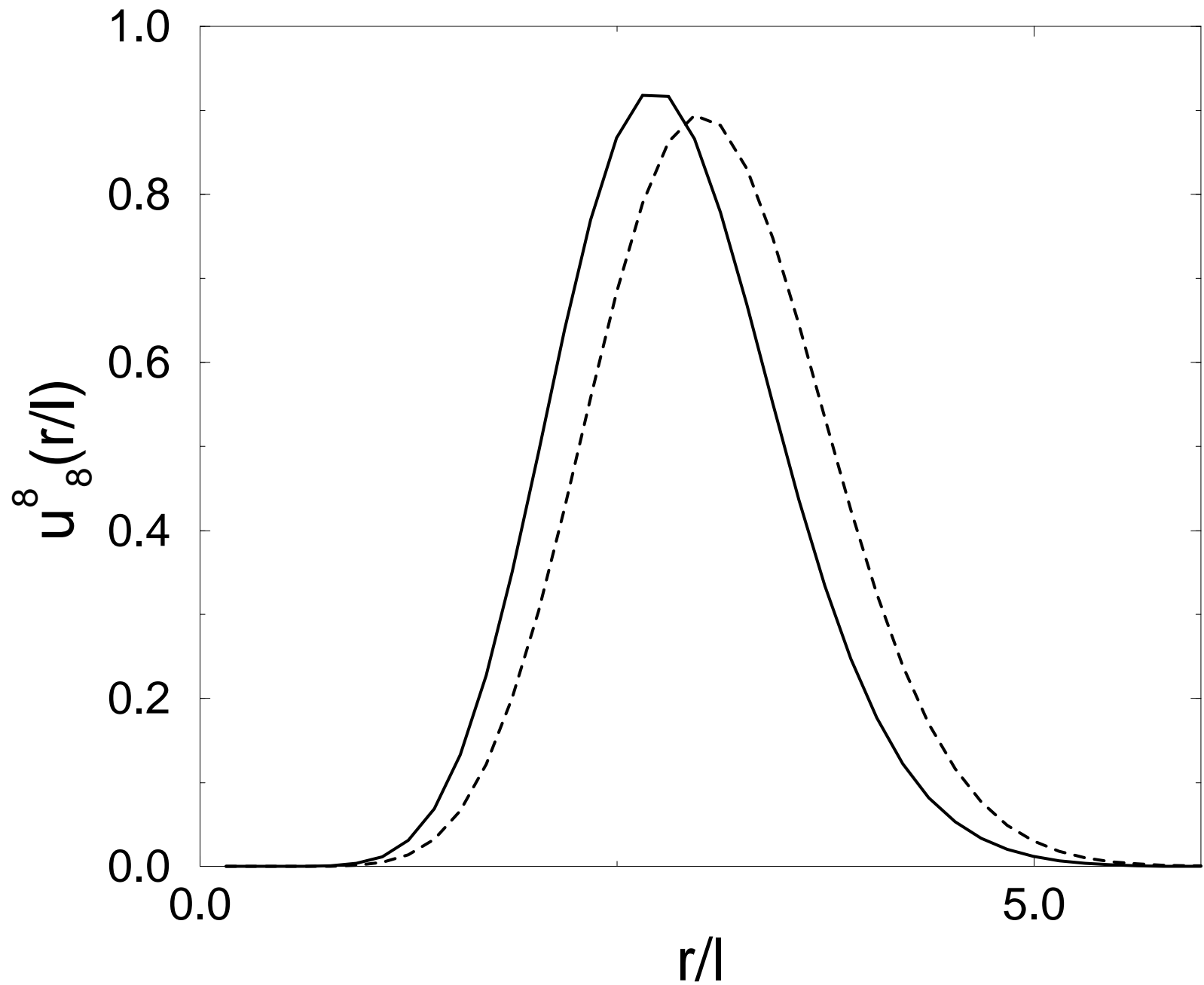
\noindent Fig.\ 3: The radial quasiparticle wave functions for $\nu=8$, $l=0$ (a) and
 $\nu=8$, $l=8$ (b) as 
a function of $r/l$. The solid lines are for $g'=2$ and the dashed lines are for $g=0$.
 The Hartree field in units of $\hbar\omega$ is plotted in (c).

\

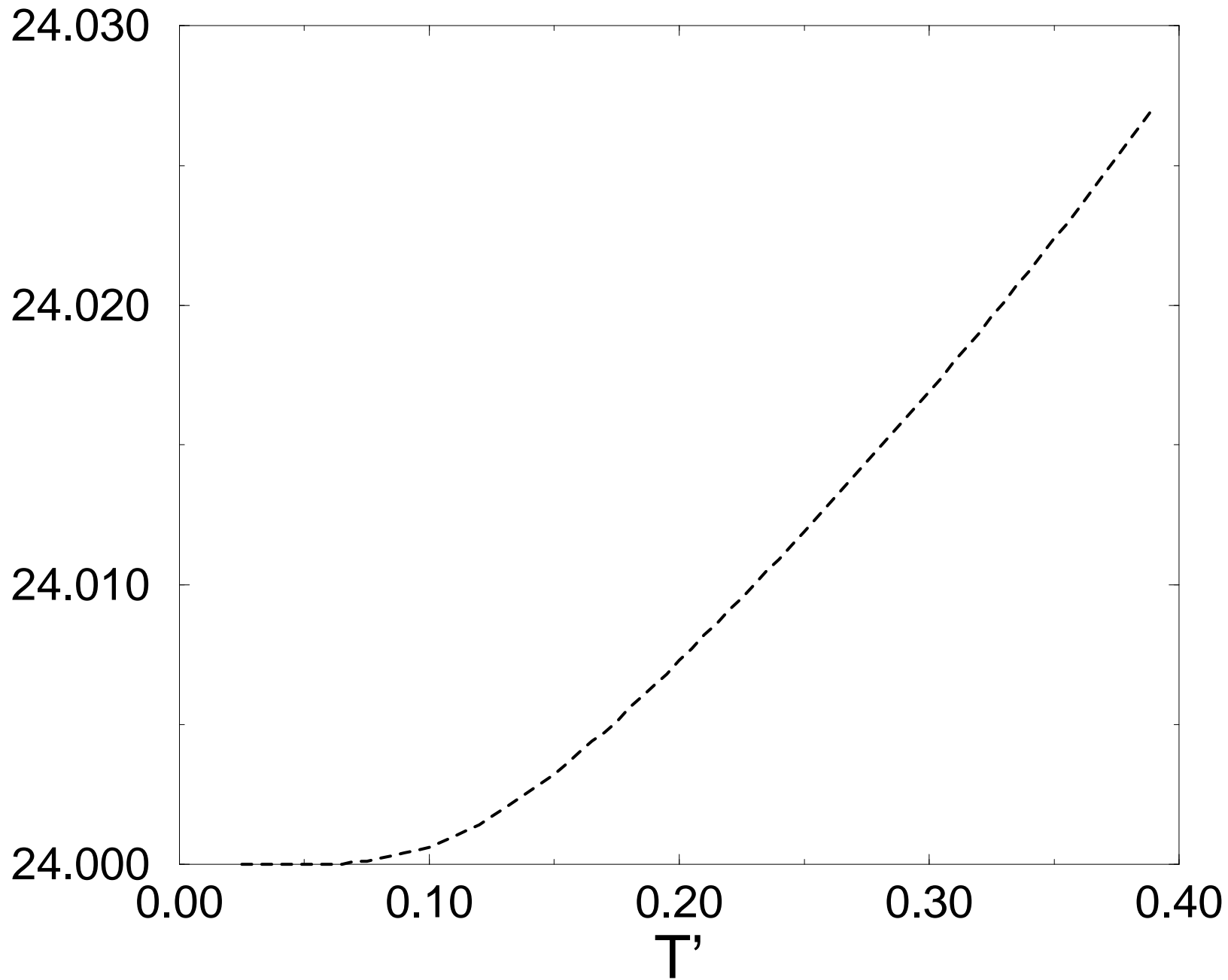
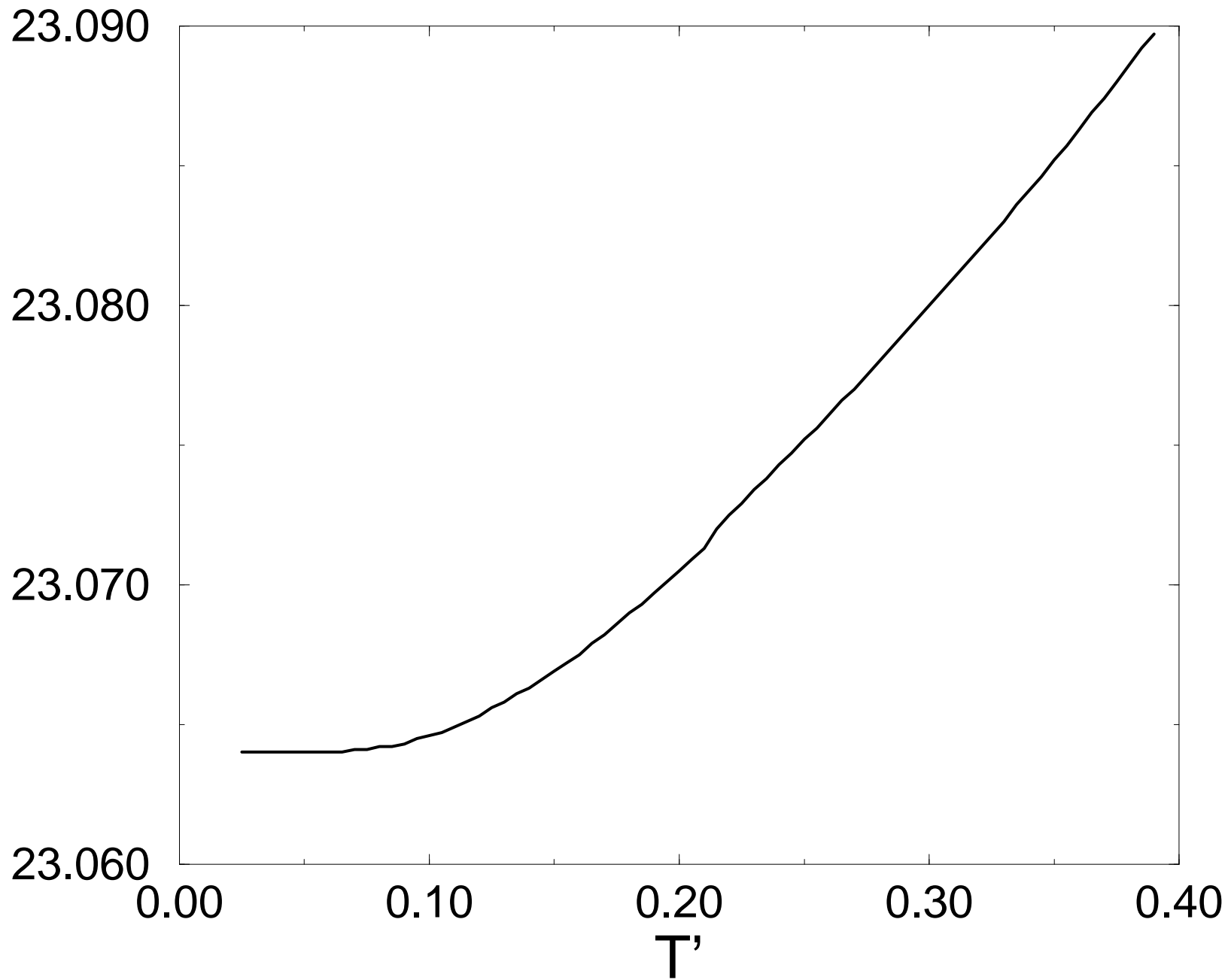
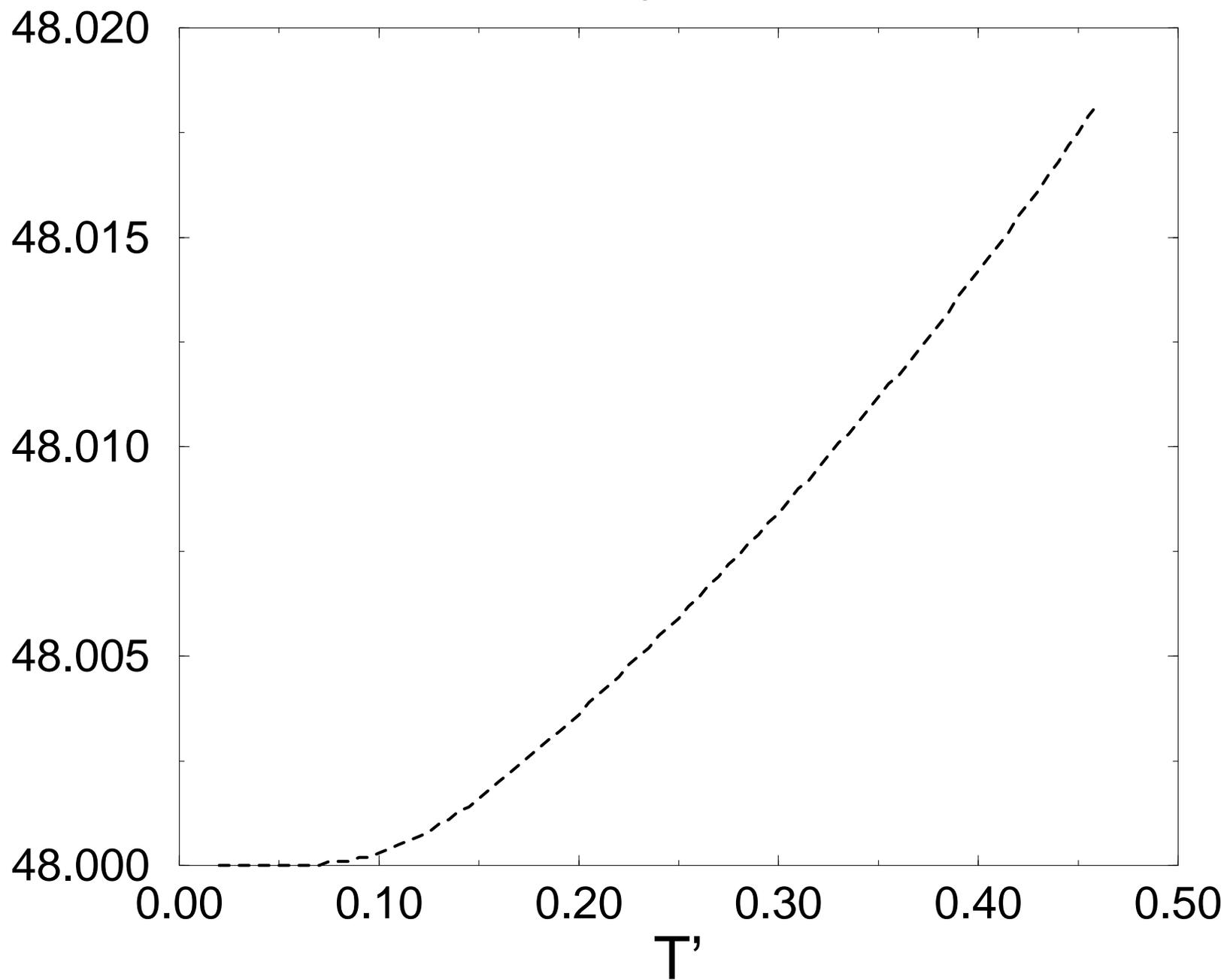
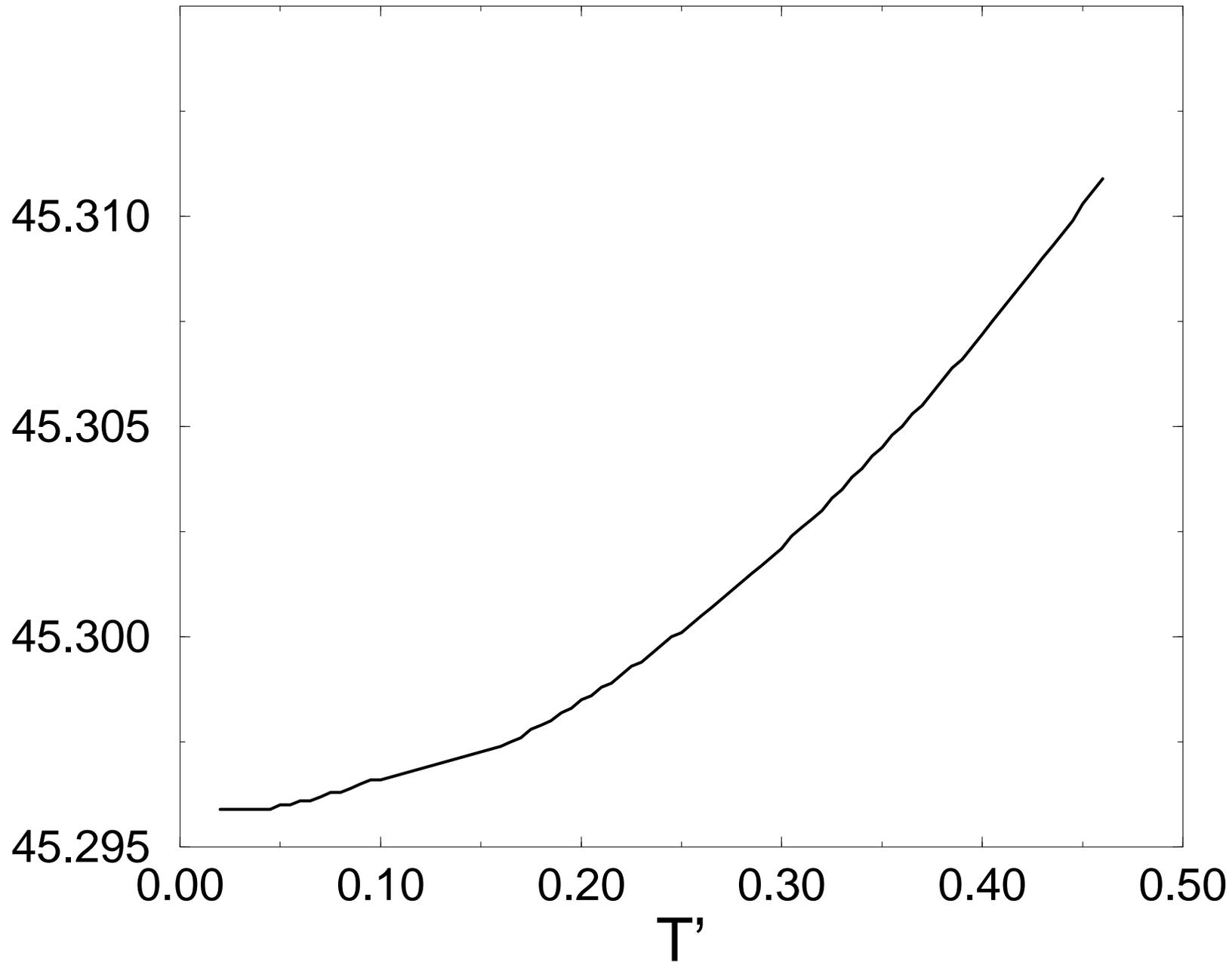
\noindent Fig.\ 4: The energy in units of $\hbar\omega$ for $g'=0.4$ (solid lines) and 
$N=5456$ (a) and 
 $N=43680$ (b). The dashed lines in are for the non-interacting case 
for $N=5456$ (c) and  $N=43680$ (d).

\

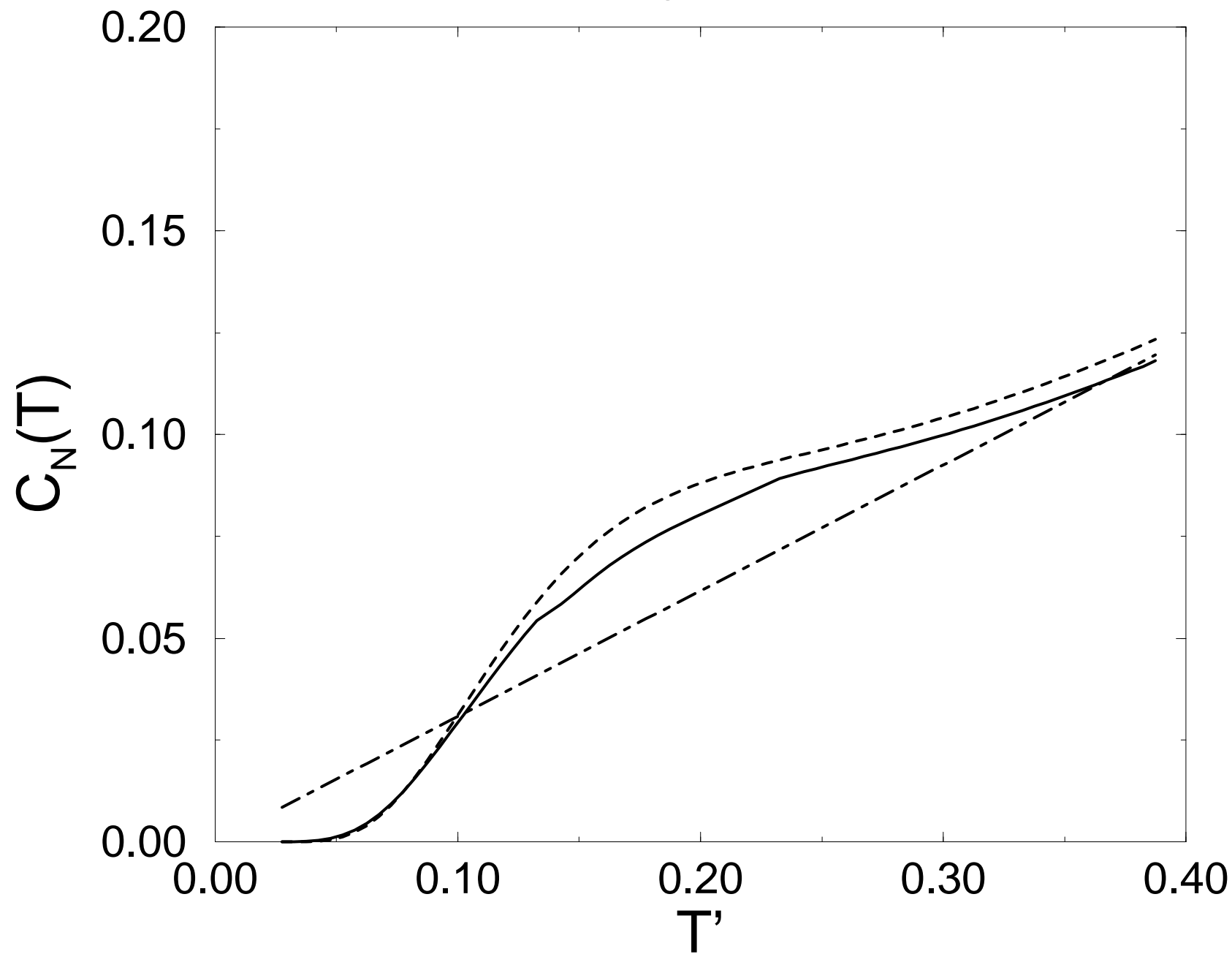
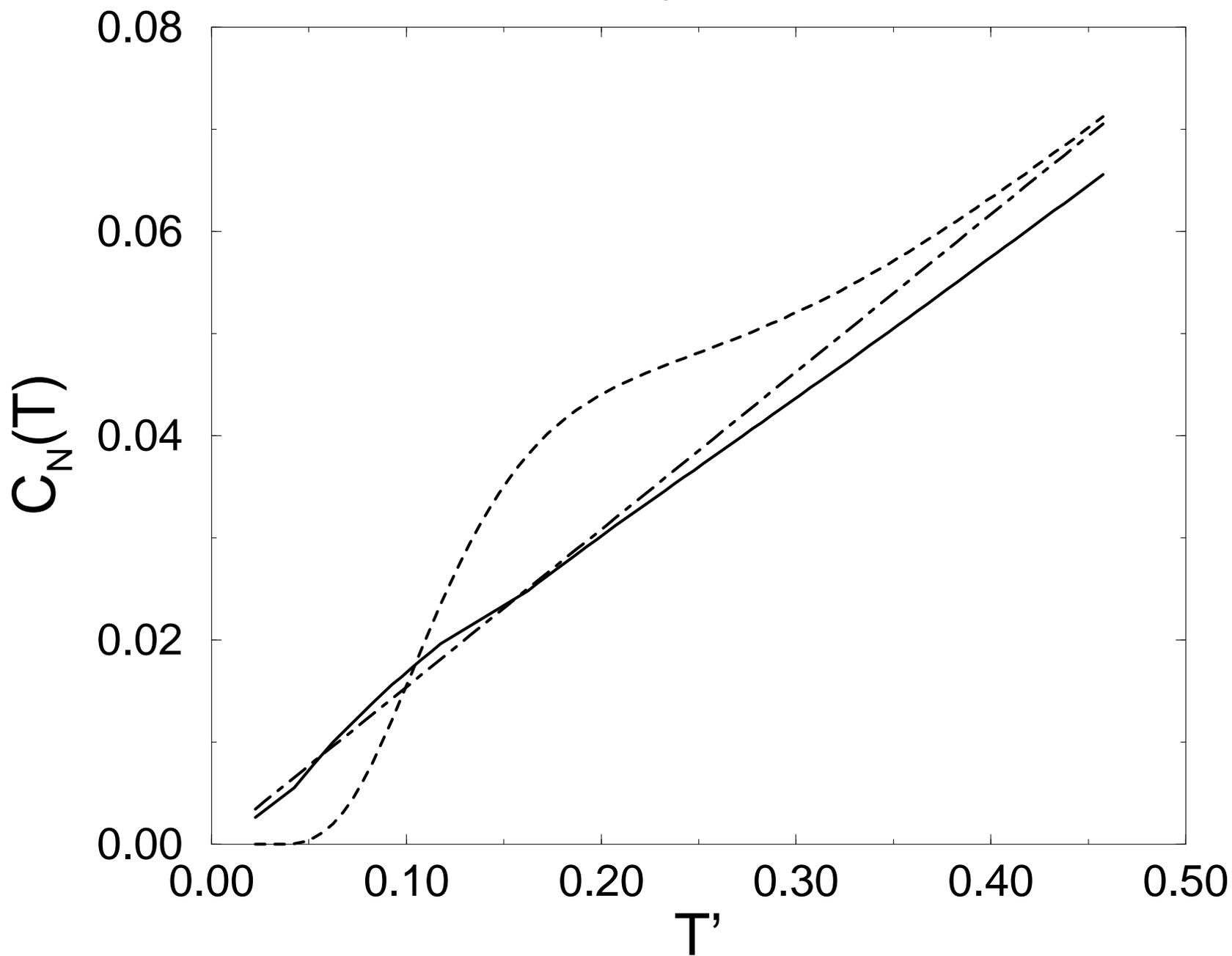
\noindent Fig.\ 5: The heat capacity  in units of $k_B$ for $g'=0.4$ and $N=5456$ (a) and 
$N=43680$ (b).

\

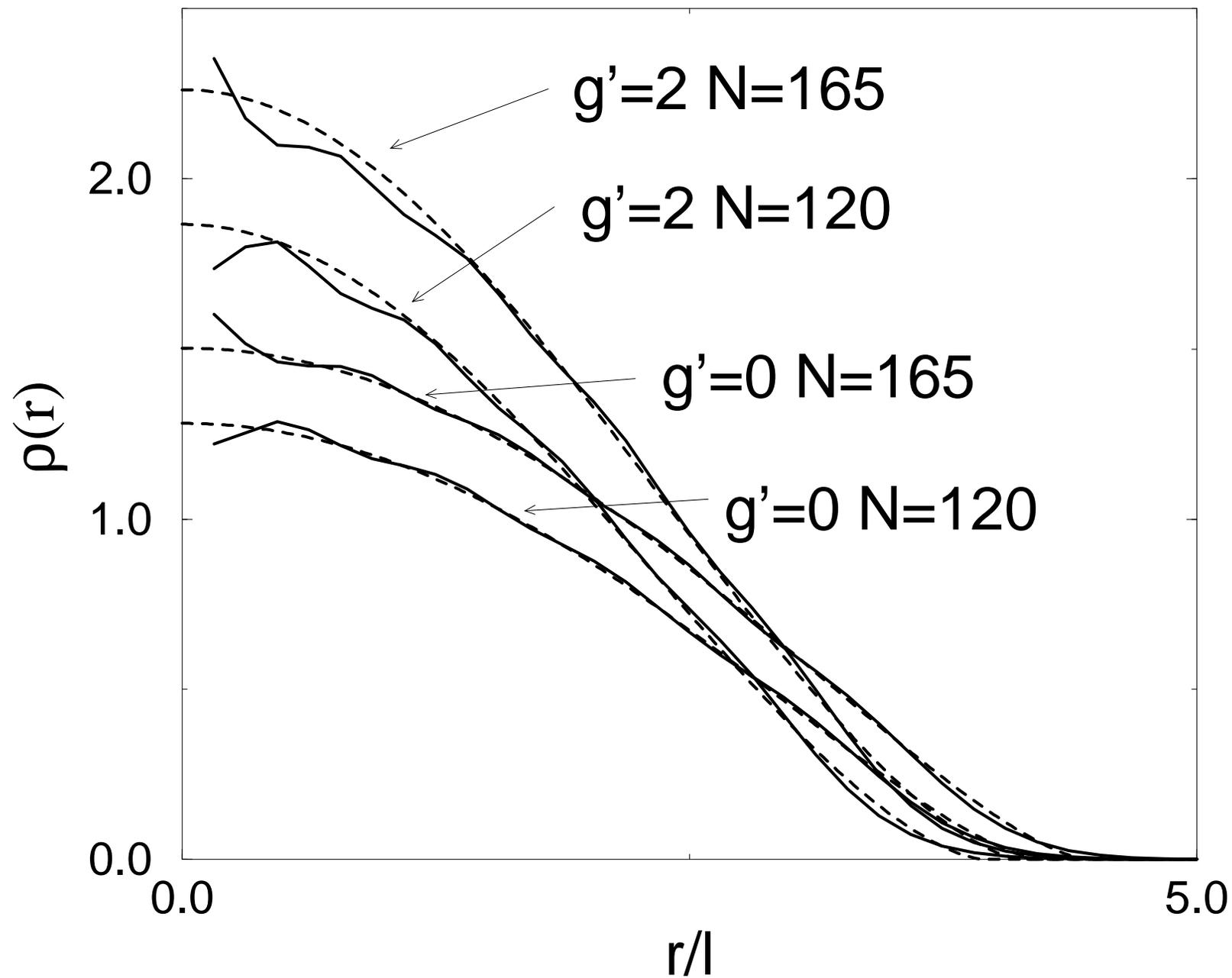
\noindent Fig.\ 6: The density distribution $\rho(r/l)$for $N=120, 165$, $g'=0, 2$ (solid lines).
The dashed lines are the Thomas-Fermi results.

\

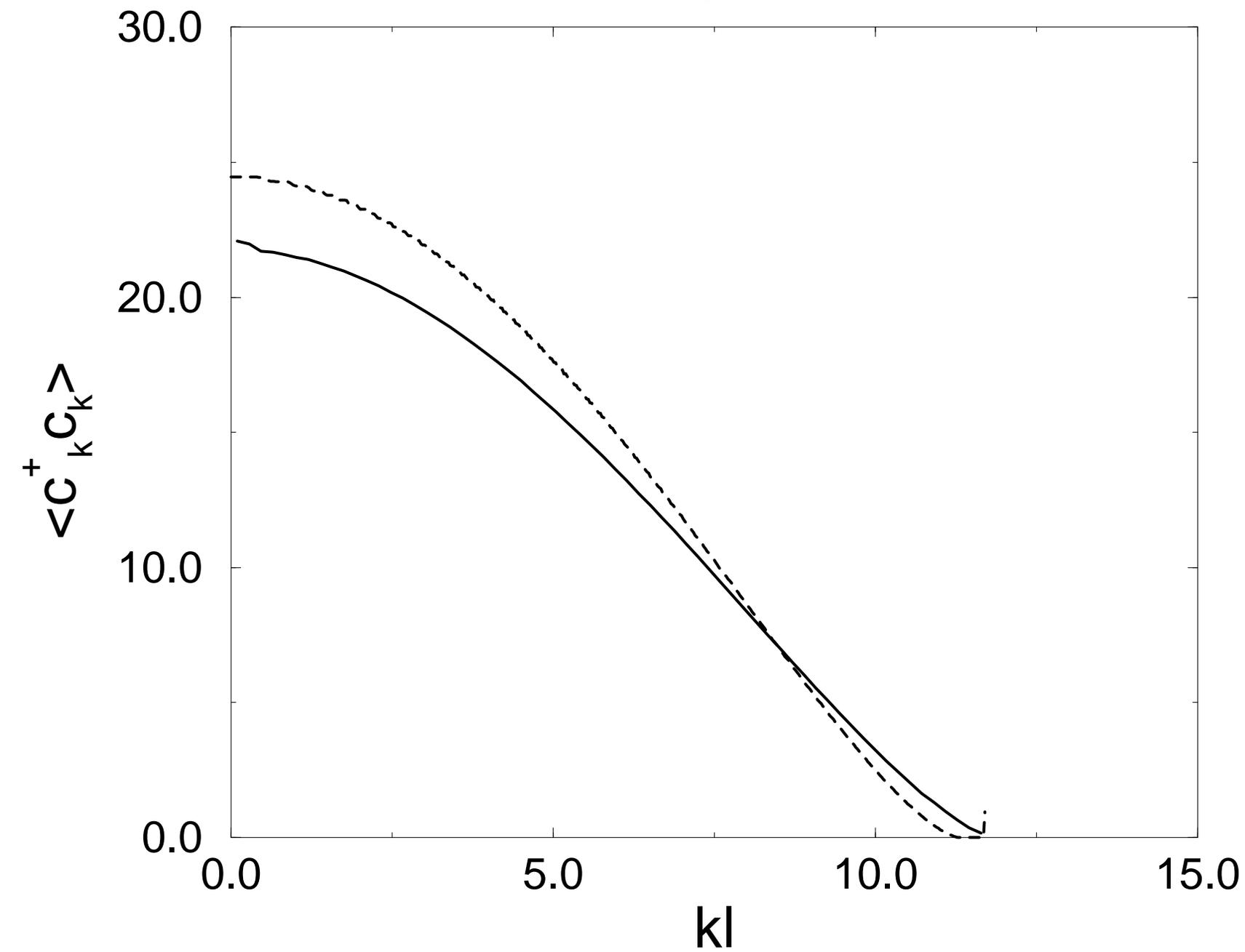
\noindent Fig.\ 7: The momentum distribution $\langle c^\dagger_kc_k\rangle$
for $N=43680$, $g'=0.4$ (solid)
and  $N=43680$, $g'=0.0$ (dashed)

\end{document}